\documentclass[11pt]{article}
\usepackage{geometry}                
\usepackage{graphicx}
\usepackage{amssymb}
\usepackage{epstopdf}
\usepackage{fmtcount}

\usepackage{amsmath, amsthm, amssymb}
\usepackage{subfigure}

\title{Power-law weighted networks from local attachments}
\author{P. Moriano and J. Finke\\ Department of Electrical Engineering and Computer Science \\ Pontificia Universidad Javeriana, Santiago de Cali\\pamoriano@javerianacali.edu.co, finke@ieee.org}
\date{}

\begin{document}

\maketitle
\pagenumbering{arabic}

\begin{abstract}

This letter introduces a mechanism for constructing, through a process of distributed decision-making, substrates for the study of collective dynamics on extended power-law weighted networks with both a desired scaling exponent and a fixed clustering coefficient. The analytical results show that the connectivity distribution converges to the scaling behavior often found in social and engineering systems. To illustrate the approach of the proposed framework we generate network substrates that resemble steady state properties of the empirical citation distributions of $(i)$ publications indexed by the Institute for Scientific Information from 1981 to 1997; $(ii)$ patents granted by the U.S. Patent and Trademark Office from 1975 to 1999; and $(iii)$ opinions written by the Supreme Court and the cases they cite from 1754 to 2002.

\end{abstract}


\section{Introduction}

Understanding structure lies at the very heart of the study of complex networks. A network is a collection of a large number of interconnected elements (units or agents) whose interaction with each other and with the surroundings leads to characteristic properties that can only be attributed to the network as a whole~\cite{Amaral:2004ph}. Networks often develop distinct structural steady state patterns. Studying these patterns, promises to enhance our understanding of the dynamics underlying collective human responses \cite{Cho:2009li}, corrupt behavior \cite{Bohannon:2009qv}, and economic development \cite{Farmer:2009kl}.

Random graph models fail to capture key features of real-world networks  (\textit{e.g.}, clustering coefficients and degree correlations). Recent efforts to understand network structure have focused on connectivity distributions underlying a number of social and engineering systems which, rather than following the Poisson distribution of random networks (bounded by Chebyshev's inequality), have heavy tails \cite{J.-Finke:2008mz}. Heavy-tailed distributions in empirical data suggests the existence of causal mechanisms that shape the structure and function of real-world networks \cite{Clauset:2009lr}. In the era of ``big data,'' the development of formal frameworks that quantify patterns of interaction of networks has set the research agendas across various disciplines (\textit{e.g.}, more recently across the data driven computational social sciences).

Power-laws, a particular type of heavy-tailed distributions, have received significant attention in recent years. For a network with an extended power-law connectivity distribution, if the number of connections of a node is much larger than $x_0$, the probability that the node connects to $x$ other nodes is proportional to $x^{-\alpha}$ for some positive constants ${\alpha}$ and $x_0$ \cite{Valverde:2007fk}. As a result, the tail of the distribution has no exponential bound and the connectivity of the nodes of the network comprises different orders of magnitude, with a few nodes being highly connected.

Key to modeling power-law networks  is the characterization of \textit{hubs} (highly interconnected nodes). In the context of the spread of disease, measuring patterns in regions that are more vulnerable to infection (hubs) allows us to respond more effectively to the potential spread of large-scale epidemics  \cite{Christakis:2010fk}. The ability to understand and recreate the structure of epidemic networks allows us to design strategies that embrace how interconnected regions influence one another (as a result of the evolution of social systems) in order to quantify and predict the dimensions of disease.

To capture the relationships between the elements of a network, \textit{e.g.}, duration, emotional intensity, or intimacy, models define \emph{weights} as an inherent property between nodes \cite{barrat:2004uq}. Recent models of weighted networks have focused on attachment strategies in which nodes are added according to probability distributions on the existing weights across the entire network. The network model introduced in \cite{Barratt:2004yq} captures the evolution of weights driven by preferential strength attachment, a mechanism in which newly added nodes are more likely to connect to nodes associated with larger weights. Lacking local competitive factors between nodes, the resulting networks exhibit power-law distributions where the hubs correspond to the nodes that have been part of the network the longest.

This letter introduces a wide class of attachment strategies which promote the formation of hubs based on both the length of time a node has been part of the network (i.e., node longevity) and its ability to compete for weights with surrounding neighbors (i.e., node fitness). Because the connectivity dynamics of the nodes depend on their attractiveness to compete for weights (as in \cite{Bianconi:2001fk}), older nodes are not necessarily more successful in acquiring weights. To our knowledge the proposed mechanism is novel in that it generates weighted directed networks with extended power-law strength distributions $(i)$ in a distributed fashion (decision-making strategies are based on local information; we do not assume any type of global information to generate the desired network structure); $(ii)$ for an arbitrary scaling exponent $\alpha > 2$ and a fixed clustering $c \in (0,1)$ (as in \cite{Holme:2002uq}); and $(iii)$ for values greater than a particular threshold $\hat{s} > 0$ (for the case when only the tail of the distribution obeys a power-law). 

The remaining sections are organized as follows. First we introduce a model that captures the connectivity and growth dynamics of the gradual addition of nodes to an existing network component and proposes attachment strategies for local rearrangement of weights between pairs of nodes. We prove that for any connected network there exists a distribution of the total weight from neighboring nodes (node strength) that is asymptotically stable (\textit{i.e.}, the proposed strategies lead to a Nash equilibrium \cite{Jackson:2005fk}). Moreover, as the network grows, consecutive achievements of this network state leads to weighted directed networks with extended power-law strength distributions and distinctive clustering coefficients (defined as the ratio of the total average weight of transitive triplets over the total weight of possible triplets). We present simulations that capture the effect of node fitness and illustrate the application of the proposed model to generate various citation networks. Finally, we draw some conclusions and future research directions.

\section{A model of network topology and growth}

Consider a directed network that captures weighted relationships between a set of nodes. As the network grows, more nodes join the network, each possessing a small ``budget'' used to construct directed links to some existing nodes. When, node $i$ establishes a link to node $j$ (by passing some of its budget to node $j$) node $j$ has more budget to spend, which it may do by increasing its weighted connections to other nodes. Broadly speaking, every node wishes to spend its budget, but the more it spends the less willing it is to spend more. Nodes will locally rearrange their weights until every node reaches an equilibrium. At the equilibrium all  nodes have associated gains that are equal and there are no further incentives to rearrange connections.

To formalize this idea let us introduce the following notation. Let $\mathcal{H}_1 = \{1, \ldots, N_1\}$ be a finite set of nodes at generation $k=1$. Nodes represent elements (acting units) that establish connections to other nodes. We represent the relationship between nodes using  a weighted matrix $\mathcal{W}_1 = [w_{ij}]_{N_1 \times N_1}$, where $w_{ij} \in \mathbb{R_+}=(0,\infty)$ quantifies the relationship between node $i$ and $j$. If $w_{ij}>0$, then there exists some kind of action from $i$ to $j$ with weight $w_{ij}$. It may capture, for instance, the extent to which node $i$ influences node $j$. Let $\mathcal{G}_k=(\mathcal{H}_k,\mathcal{W}_k)$ represent the network at generation $k$ (because in general $w_{ij} \ne w_{ji}$, the network is modeled as a directed graph). For a fixed generation, let $p(i) = \{j:w_{ji}>0\}$ represent all nodes which influence node $i$ (incoming neighbors). Similarly, let $q(i) = \{j:w_{ij}>0\}$ represent all nodes influenced by node $i$ (outgoing neighbors). A gain function $g_{i}(s_{i})$ is associated to each node $i \in \mathcal{H}_k$ and characterizes the marginal benefit that results from its current set of connections, where $s_{i} = \sum_{j \in p(i)} w_{ji}$, $s_{i} \in \mathbb{R}_{+}$. Note that $s_i$ is a scalar that represents the \emph{incoming strength} of node $i$ (referred to as node strength hereafter). The following network assumptions are needed:

\begin{itemize}
 
\item [A1] \emph{Finite network strength:} The total weight of the initial network $P_1=\sum_{i=1}^{n_1}s_{i}$, $P_1 \in \mathbb{R}_{+}$, is finite. In other words, the extent to which any node in the network can be influenced by other nodes is bounded.

\item [A2]  \emph{Connectedness:} Every node is influenced to some extent by another node. At each generation $k$, $s_i \ge \epsilon > 0, \ \forall i \in \mathcal{H}_k$.

\item [A3] {\itshape Bounded marginal gains:} The gain function $g_i(s_i)>0$ associated to node $i \in \mathcal{H}_{k}$ satisfies
\begin{eqnarray} \label{e1}
-a_i \le \frac{g_i(y_i)-g_i(z_i)}{y_i-z_i} \le -b_i
\end{eqnarray}
for any $y_i, z_i \in \mathbb{R}_{+}$, $y_i \ne z_i$ and some constants $a_i \ge b_i > 0$. In other words, the marginal gain associated with each node decreases with increasing strength. Equation~\eqref{e1} eliminates the possibility that a very small difference in node strength may result in an unbounded change in gain. Note that if $g_i$ is differentiable and has a negative derivative it satisfies eq.~\eqref{e1}. 

\end{itemize} 


Next, we use $t \ge 0$ to specify the time index of events. Let $t=\tau_k$ be the time instant when a new node is added to form the network $\mathcal{G}_k$ (\textit{i.e.}, the start of generation $k$). Let $\tau^+_{k}$ be the instant right before the new node is added to $\mathcal{G}_k$ (\textit{i.e.}, the start of generation $k+1$). When $t = \tau_{k+1}$, $\mathcal{G}_k$ evolves into $\mathcal{G}_{k+1}$. For generation $k$ let the set of states
\begin{eqnarray*} \label{e2}
\mathcal{S}_k = \left \{ s \in \mathbb{R}_{+}^{N_k}: \sum_{i=1}^{N_k}s_{i}=P_k \right \}
\end{eqnarray*}
be the simplex over which the connectivity dynamics evolve. Constraints on our model below will ensure that for all nodes $i \in \mathcal{H}_k$, $s(t) \in \mathcal{S}_k$ for all $\tau_k \le t < \tau_{k+1}$. We assume that as $t \to \tau^+_k$, the time allowed for the events that drive the connectivity dynamics during generation $k$ goes to infinity. Let $s(t)=[{s_1}(t),\ldots, s_{N_k}(t)]^\top \in \mathcal{S}_k$ be the state vector for $\mathcal{G}_k$ at time $t$ (\textit{i.e.}, the incoming strength distribution of the entire network). 

\subsection{Connectivity dynamics}

We first focus on the dynamics of $s(t)$ for $\tau_{k} \le t < \tau_{k+1}$ (\textit{i.e.}, within a fixed generation). In particular, we want to define the singleton
\begin{eqnarray} \label{e3}
\mathcal{S}^*_k=\{s \in \mathcal{S}_k: \mbox{for all} \; i, j \in \mathcal{H}_k, \; g_i(s_i)=g_j(s_j) \} 
\end{eqnarray}
such that any strength distribution that belongs to this set represents a distribution where all nodes in $\mathcal{H}_k$ have equal gain levels. To capture the connectivity dynamics that lead to $\mathcal{S}_k^{*}$, let $e_{\mu(i)}^{\sigma(i)}$ represents the decision of node $i$ to weaken its relation from some nodes $j$ in $p(i)$ while strengthening its relation to other nodes in $\mathcal{H}_k$. Let the list $\sigma(i)=(\sigma_j(i),\sigma_{j^{'}}(i),\ldots,\sigma_{j^{''}}(i))$ such that $j<j^{'}<\cdots<j^{''}$ and $j,j^{'},\ldots,j^{''} \in \mathcal{H}_k$ be composed of elements $\sigma_{j}(i)$ that denote the weight to be added or created to the to link $\omega_{ij}$ between node $i$ and node $j \in \mathcal{H}_k$. For convenience, we will denote this list by $\sigma(i)=(\sigma_{j}(i):j \in \mathcal{H}_k)$. Similarly, let the list $(\mu_j (i): j \in p(i))$ be composed of elements $\mu_{j}(i)$ that denote the weight to be subtracted from the link $w_{ji}$ where node $j \in p(i)$. 

Let $\{ e_{\mu(i)}^{\sigma(i)} \}$ denote the set of {\itshape all} possible combinations of how node $i$ can weaken or strengthen its relations to other nodes. Let the set of events be described by $ \mathcal{E}_1= \mathcal{P} \left (\{ e_{\mu(i)}^{\sigma(i)} \} \right)  -\{ \emptyset \}$ ($\mathcal{P}$($\cdot$) denotes the power set). We call $e_1 (t)$, $\tau_k \le t < \tau_{k+1}$, events of type $1$; they drive the connectivity dynamics within a network generation. Notice that each event $e_1(t) \in \mathcal{E}_1$ is defined as a {\itshape set}, with each element of $e_1(t)$ representing the potential rearrangement of multiple weights between nodes, and multiple elements in $e_1(t)$ representing the simultaneous rearrangements among multiple nodes.

An event $e_1(t)$ may occur only if it belongs to the set defined by an enable function $h_{1}:\mathcal{S}_k \longrightarrow \mathcal{P}(\mathcal{E}_1)-\{ \emptyset \}$, specified for node $i \in \mathcal{H}_k$ as follows 
\begin{itemize}
\item[-] If $g_i(s_i) \ge g_j(s_j)$ for all $j \in {q(i)}$, then $e_{\mu(i)}^{\sigma(i)} \in e_1(t)$ such that $\sigma(i)=(0, \ldots,0)$ and $\mu(i)=(0, \ldots,0)$ is the only enabled event. Hence, node $i$ does not modify its relationships to others nodes (\textit{i.e.}, the strength of node $i$ does not change).
\item[-] If $g_i(s_i) < g_j(s_j)$ for some $j \in q(i)$, then the only  $ e_{\mu(i)}^{\sigma(i)}  \in e_1(t)$ are ones with $\sigma(i)=(\sigma_j(i):j \in \mathcal{H}_k)$ and $\mu(i)=(\mu_j(i):j \in p(i))$ such that
\begin{eqnarray*} \label{e4}
\mbox{C1}  && \sum_{j \in \mathcal{H}_k} \sigma_j(i)=\sum_{j \in p(i)} \mu_j(i) \\
\mbox{C2}  && \sigma_{j^*}(i) \ge \frac{1}{a_i}{\gamma \left(g_{j^*}(s_{j^*})-g_i(s_i)\right)} \\
\mbox{C3}  && \sum_{j \in p(i)}\mu_j (i) \le \frac{1}{b_i}\left(g_{j^*}(s_{j^*})-g_i(s_i)\right)-\sigma_{j^*}(i)
\end{eqnarray*} 
\end{itemize} 
for some $j^* \in \{ j:\; g_j(s_j) \geq g_r(s_r)$, for all $r \in q(i)\}$ and $\gamma$. The parameter~$\gamma \in (0,1)$ regulates the speed at which weights are rearranged and affects the transitivity of the network (\textit{i.e.}, if a node $j$ is connected to node $j'$ and node $j'$ to node $j''$, the probability that node $j$ is also connected to node $j''$). Low values of $\gamma$ lead to slower convergence processes which increase the probability of forming transitive triples and lead to higher clustering coefficients.  

Condition C1 implies that a node can only establish or strengthen its relations to other nodes by weakening incoming weights (the sum of incoming weights must equal the sum of outgoing weights). It implies that $\mathcal{G}_k$ conserves total network strength, \textit{i.e.}, $P_k=\sum_{i=1}^{N_k}s_{i}(t)$ is constant. To interpret C2 and C3 it is useful to remember that reducing (increasing) the strength of a node always increases (decreases, respectively) its gain. Both conditions constrain how nodes can modify their weights in terms of the gain of outgoing neighbors. Condition C2 implies that if the gain of node $i$ differs from any of its outgoing neighbors, then the relation to some neighbor with the highest gain must be strengthened by some amount. Condition C3 implies that when node $i$ weakens incoming weights, node $i$ cannot exceed the highest gain of at least one outgoing neighbor. Together they guarantee that the highest gain of the network is strictly monotonically decreasing over time (as we prove in Theorem 1).

Next, state transitions are defined by the operator $f_{1}: \mathcal{S}_k \longrightarrow \mathcal{S}_k$ where $e_1(t) \in \mathcal{E}_1$. For a fixed generation $k$, if $e_1(t) \in h_{1}(s(t))$, $e_{\mu(i)}^{\sigma(i)} \in e_1(t)$, then $s(t+1)=f_{1}(s(t))$, where 
\begin{eqnarray} \label{e5}
s_i(t+1)&=&s_i(t) \ +\sum_{\{j \; \in \;\mathcal{H}_k,\;e_{\mu(j)}^{\sigma(j)} \in \; e_1 (t)\}} \sigma_{i}(j) \nonumber \\
&-&\sum_{\{j\; \in \; p(i),\;e_{\mu(i)}^{\sigma(i)} \in \; e_1 (t)\}} \mu_{j}(i)
\end{eqnarray}
Equation~\eqref{e5} means that the strength at node $i$ at time $t+1$ equals the strength of node $i$ at time $t$, plus the total weight added by the nodes that strengthened their relationship to node $i$, minus the total weight reduced by nodes that weakened their relation to node $i$ at time $t$.

Let $E_1$ denote the set of all infinite sequence of events $\mathcal{E}_1$. Let $E^{1}_t$ denote the sequence of events $e_1(0), \ldots, e_1(t-1)$ and let the value of the function $S(s(0),E^{1}_t,t)$ denote the state reached at time $t$ from the initial state $s(0)$ by the application of the sequence $E^{1}_t$ of events of type 1. We assume that each event of type 1 occurs infinitely often on each event trajectory $E^{1}_tE^{1}$, $\tau_k \le t < \tau_{k+1}$. This assumption is met if nodes persistently try to rearrange weights. The enable function $h_{1}$ together with state transition operator $f_{1}$ define the evolution of the connectivity dynamics of the network. 

\subsection{Growth dynamics}

We now turn our attention to the evolution of the network as it grows. To capture a nodes's advantage of longevity let $k_i$ be the generation when node $i$ is added and define $n_i=\frac{k_i}{k}$ as the fraction of generations node $i$ has \emph{not} been part of the network component. Moreover, to capture a node's competitive advantage in acquiring weights we associate to every node a fitness $\beta_i$, where $\beta_i \in (0,1)$. Let $s_0 \ge 0$ be a constant amount of strength such that $s_i > s_0$. Let the gain function (marginal utility) associated to node $i \in \mathcal{H}_k$ during generation $k$ be
\begin{eqnarray} \label{e6}
g_i(s_{i})=\frac{1}{s_i-s_0}\left(\frac{1}{n_i} \right)^{\beta_i}
\end{eqnarray}
Higher values of $\beta_i$ characterize nodes that are more attractive in the sense that they can carry more weight without greatly reducing their gain. Both high values of $n_i$ (representing the fact that node $i$ has been part of the growing network for only a few generations) and low values of $\beta_i$ (representing the fact that the node has a low competitive advantage for acquiring weights) have a negative effect on the gain of node $i$. Below we will see how $\beta_i$ allows us to define the scaling exponent of extended power-law strength distributions. 

Let $e^{\sigma(i)}$ represents the attachment of a new node $i$ to the network at the beginning of generation $k$ (when $t=\tau_{k}$). Let $m=\sum_{r} \sigma_r (i)$ be the total (constant) weight of a newly added node. A node attaches to the network component by $(i)$ randomly distributing its weight $\sigma(i)$ across some nodes and $(ii)$ establishing a non-empty set of incoming neighbors (\textit{i.e.}, some node must connect to it). We call the attachment of nodes to $\mathcal{G}_k$, $k=1,2,\ldots$ events of type~2. Let $\mathcal{E}_2 =\{e^{\sigma(i)}\}$ denote all possible combinations of how node $i$ can attach to the network component. An event $e_2 (k) \in \mathcal{E}_2$ may occur if it is defined by an enable function~$h_{2}: \mathcal{S}_k \longrightarrow e^{\sigma(i)}$, specified for a newly added node as follows
\begin{itemize}
\item[-] Node $i$ attaches to the network only if the associated gain function $g_i(s_i)$ follows the general form of \eqref{e6} with longevity and fitness parameters that satisfy
\end{itemize}
\begin{eqnarray*}
\mbox{C4} && n_{i}=1 \\
\mbox{C5} && \beta_i = \beta \ \forall i \in \mathcal{H}_k
\end{eqnarray*}
Condition C4 follows from letting $k_{i} = k$ for the newly added node (at generation $k$ node $i$ has been part of network for one generation). Condition C5 specifies an equal fitness value for every node (as is the case for networks with linear growth under preferential attachment).

The transition $e_2 (k) \in \mathcal{E}_2$ is defined by the operator $f_{2}:~ \mathcal{S}^*_k  \longrightarrow \mathcal{S}_{k+1} $. If $e_{2}(k) \in h_{2} (s(\tau_k))$, then $s(\tau_{k+1})=f_{2}(s(\tau^+_k))$ where $s_{i}(\tau_{k+1})=m$ only if node $i$ is the newly added node. Let $E_2$ denote the set of all infinite sequence of events $\mathcal{E}_2$. Let $E^{2}_k$ denote the sequence of events of type 2, $e_2(1), \ldots, e_2(k)$. We assume that each event of type 2 occurs infinitely often on each event trajectory $E^{2}_kE^{2}$. The assumption is met if nodes constantly attach to the existing network component. The enable function $h_{2}$ together with the transition operator $f_{2}$ define the growth dynamics of the network. 

\section{Analysis}

Next, we present stability properties of the invariant set $\mathcal{S}^*_k$ and deduce the average gain level of the network $\mathcal{G}_k$. We then prove that, for values greater than a threshold $\hat{s}$, the strength distribution converges to a scaling behavior.

\noindent \emph{Theorem $1$: Suppose \emph{A1-3} and \emph{C1-3} hold. Then $\mathcal{S}_k^*$ is an invariant set and has region of asymptotic stability equal to $\mathcal{S}_k$.}

Theorem~1 guarantees that for any generation $k$, initial network state $s(0)$, and event sequence $E^1_t$, $S(s(0),E^1_t ,t) \to \mathcal{S}^*_k$ as $t \to \tau_{k+1}$ for generation $k$. Broadly speaking, the conditions in Theorem 1 capture the dynamic coupling between different nodes that lead to a Nash equilibrium. By attaining the same gain level no node can increase its gain by changing its connections unilaterally without making the average gain of all other nodes worse off. When $\mathcal{S}^*_k$ is reached the average gain of the network an instant before the start of generation $k+1$ is given by
\begin{eqnarray}\label{e50}
C_k=\frac{1}{N_1 +k}\sum_{i \in \mathcal{H}_k}g_i (s_i (\tau^{+}_{k}))
\end{eqnarray}
As the network grows, the behavior of the average gain is characterized by the following lemma.

\noindent  \emph{Lemma $1$: Suppose \emph{A1-3} and \emph{C1-5} hold. Moreover, $\forall k$ let $s(\tau_{k} ^{+}) \in S_k^*$ then~$C_k \to 1/(m+s_0) (1-\beta)$ as $k \to \infty$.}

Lemma 1 implies that at the desired strength distribution $\mathcal{S}^*_k$, the average gain tends to $C_k \to 1/(m+s_0) (1-\beta)$ as $k \to \infty$. 

The following theorem implies that as the network grows, it develops an extended power-law structure driven by the marginal benefit of the allocation of weights across nodes and quantifies the value $\hat{s}$ above which the scaling behavior emerges.

\noindent  \emph{Theorem $2$: Suppose \emph{A1-3} and \emph{C1-5} hold.  Moreover, $\forall k$ let $s(\tau_{k}^{+}) \in S_k^* $. Then the strength distribution $P[s_i - s_0 > \omega]$ of the network $\mathcal{G}_k(\mathcal{H}_k,\mathcal{W}_k)$ follows an extended power-law with scaling exponent $\alpha=1 / \beta + 1$ as $k \to \infty$. The scaling behavior holds for values greater than $\hat{s}=(m+s_0)(1-\beta)$.}

Note that if $s_0 = 0$ the model yields power-law rather than extended power-law distributions (as in preferential attachment with linear growth) for values greater than $m(1-\beta)$ \cite{Barratt:2004yq}. 

Extended power-law distributions emerge as a result of both the interaction between local mechanisms that lead to Nash equilibria and the continuum attachment of new nodes to the network. In particular, when the network is at a Nash and a new node is added, it introduces a perturbation to the existing set of strategies. Conditions C1-3 force the network to return to a state which again represents a Nash, with subsequent achievements of Nash equilibria shaping the structure of the network.

\section{Simulations}

\begin{figure}[htp]
\begin{center}	
\includegraphics [width=8.0cm]{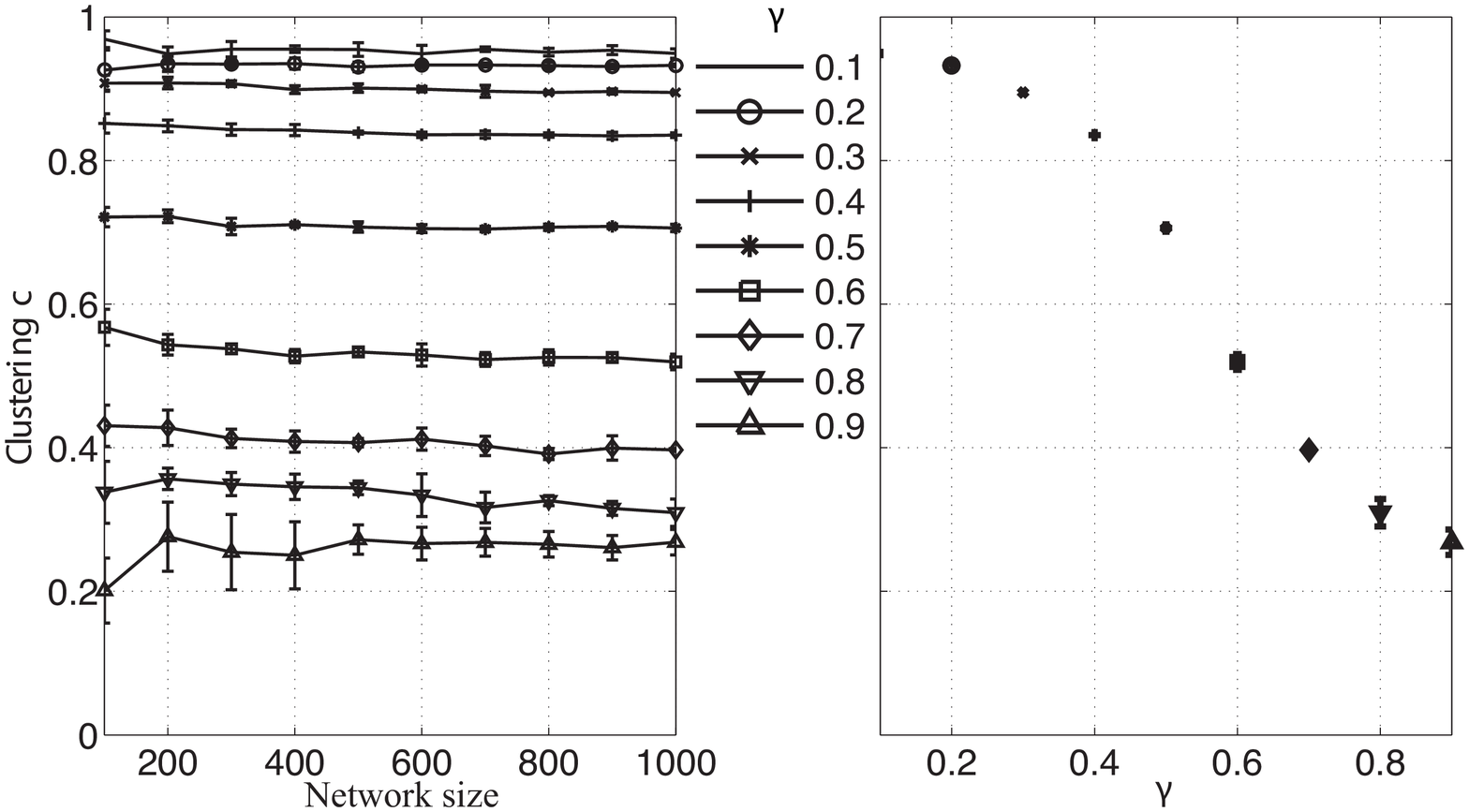}
\end{center}
\caption{Clustering coefficient as a function of network size $N_k$ at various values of $\gamma$ and as a function of $\gamma$.} 
\label{fig02}
\end{figure}

\begin{figure}[htp]
\begin{center}	
\includegraphics [width=8.0cm]{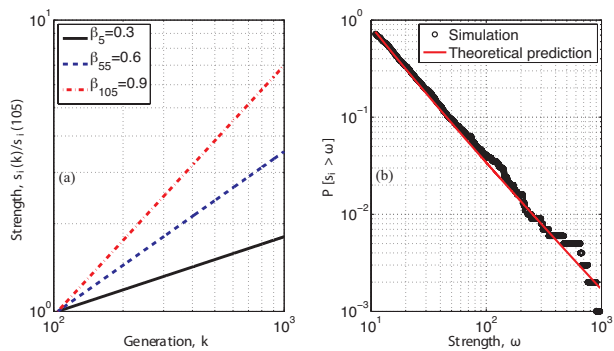}
\end{center}
\caption{(a) Evolution on the strength of three nodes added to the network $\mathcal{G}_{1000}$ using fitness $\beta_{5}=0.3$, $\beta_{55}=0.6$, and $\beta_{105}=0.9$ from $\beta_i \sim U(0,1)$ with $m=30$ and $s_0 = 0$ when $a = 1$. (b) Cumulative strength distribution~$P[s_i > \omega]\sim Ei(-C^* \log(\omega))$ or $p_{\omega} \sim \frac{1}{\log(\omega)}\omega^{-(1+C^*)}$ where $Ei(x)$ is the exponential integral function (\textit{i.e.}, a power-law with an inverse logarithmic correction term emerges).} 
\label{fig03}
\end{figure}

\begin{figure*}[htp]
\begin{center}	
\includegraphics [width=13cm]{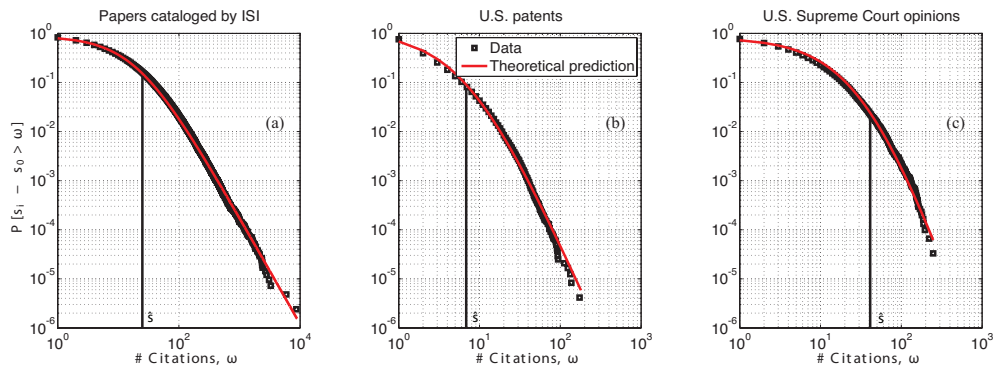} 
\end{center}
\caption{Cumulative probability distribution $P[s_i - s_0> \omega]$ for (a) the paper citation network presented in \cite{Clauset:2009lr}; (b) the U.S. patent citation network presented in \cite{Hall:2001uq}; and the U.S. Supreme Court citation network presented in \cite{Fowler:2008fk}.} 
\label{fig06}
\end{figure*}

To gain insight into the connectivity dynamics let~$\beta=\frac{1}{2}$,~$m=1$, $s_0 = 0$, $N_1=2$, and consider a network after $k=1000$ generations. Figure~\ref{fig02} shows the value of the clustering coefficient $c=\sum_{\tau_{\Delta}}\omega/\sum_{\tau}\omega$ (\textit{i.e.}, the ratio of the total average weight of transitive triplets over the total weight of possible triplets) as a function of the size. Note that for any $\gamma \in (0,1)$ the clustering properties remain constant as the network grows. 
Figure~\ref{fig03} shows the effect of varying node fitness, where $\beta_i$ is chosen from a uniform distribution with support $(0,1)$. Figure~\ref{fig03}a shows the evolution of the node's strength for different values of $\beta_i$. Note that $s_i (\tau^{+}_{k})$ follows a power-law for all values of $\beta_i \sim U(0,1)$. Because of their competitive advantage, there are some nodes with more strength $s_i$ which have been part of the network for only a few generations. It is possible for a node to join the network at a more recent generation and become more attractive than other nodes that have been part of the network for longer. In particular, fig.~\ref{fig03}a shows that the node added at generation $k=105$ with $\beta_{105} =0.9$ overcomes older nodes with $\beta_{55} =0.6$ and $\beta_{5}=0.3$. In fig.~\ref{fig03}b, the cumulative strength distribution for the entire network suggests a power-law with a logarithmic corrective term similar to the theoretical prediction in \cite{Bianconi:2001fk} where $p_{\omega} \sim \frac{1}{\log(\omega)}\omega^{-(1+C^*)}$ with $C^* = 1.255$. 
Finally, fig.~\ref{fig06} shows empirical data on the citation distribution of articles indexed by the Institute for Scientific Information (ISI); patents granted by the U.S. Patents and Trade Office; and opinions written by the U.S. Supreme Court and the cases they cite. Figure~\ref{fig06}a illustrates the case for scientific papers published in 1981 and cited between 1981 and 1997 \cite{Redner:1998lr}. The authors of \cite{Clauset:2009lr} estimated both the scaling exponent $\alpha^*=3.16$ and the threshold $\hat{s}^*=160 \pm 35$ at which the scaling behavior emerges. Figure~\ref{fig06}b represents citations on the main subnetwork of U.S. patents granted between 1963 and 1999 and references made to these patents between 1975 and 1999 \cite{Hall:2001uq}. Figure~\ref{fig06}c shows the majority opinions written by the U.S. Supreme Court and the cases they cite from 1754 to 2002 \cite{Fowler:2008fk}. All three citation networks follow extended power-law distributions (for the last two examples we estimate the values of $\alpha^*$, $\hat{s}^*$, and $c^*$ from empirical data). 
\addtocounter{footnote}{1}
\footnotetext[\value{footnote}]{For the paper citation network we use the data and the distribution predicted by the model introduced in \cite{Clauset:2009lr}. For the U.S. patent citation network we use the data presented in \cite{Hall:2001uq} and the distribution predicted by the model introduced in  \cite{Valverde:2007fk} with $\alpha=4.9$ and $\hat{s}=9.6$. For the U.S. Supreme Court citation network we use the data presented in \cite{Fowler:2008fk} and the distribution predicted by the model introduced in  \cite{Valverde:2007fk} with $\alpha=4.1$ and $\hat{s}=35$ respectively. }
\begin{table}[ht] 
\caption{Model parameters for the three citation networks.} 
\centering 
\scriptsize
\tabcolsep=0.07cm
\begin{tabular}{ c | c l c  c  c } 
\hline
& & & Papers & Patents & Court cases  \\ 
\hline 
 & Nodes & & 783339 & 240547 & 30288 \\
Empirical & Links & & 6716198 & 561060 & 220500 \\
network & $\alpha^*$& & 3.16 & 4.68 & 4.29 \\
 & $\hat{s}^*$& &160 $\pm$ 35& 8 $\pm$ 2 & 55 $\pm$ 20 \\ 
&$c^*$& &  Not available & 0.037 &0.107 \\ 
\hline 
Generated & $\alpha$& &3.13 & 4.63 & 4.25 \\
network & $\hat{s}$ &&13.3 & 7.2 & 40.8 \\ 
model &$c$& & 0.713 & 0.044 &0.112 \\ 
\hline 
 & $m$& &7 & 2.9 & 7 \\
Model & $s_0$ &&18 & 7 & 44 \\ 
parameters&$\beta$& & 0.47 & 0.28 &0.31 \\ 
&$\gamma$& & 0.5 & 0.98 &0.93 \\ 
\hline
D-statistic & Proposed model & &0.2792 & 0.1656 & 0.1761 \\
 & Previous models$^{\decimal{footnote}}$ & &  0.9910 &  0.3681 &  0.2995 \\
\hline
Sum of squares & Proposed model & &0.7358 & 0.0599 & 0.0499 \\
& Previous models$^{\decimal{footnote}}$ & &  2.3829 & 0.1416 & 0.1803 \\
\hline
Range & & & [160, 8904] & [0, 173]  & [0, 248]  \\
\hline
\end{tabular} 
\label{table1} 
\end{table}

Finally, we compare the distributions from empirical data with the distributions predicted by the proposed and previous models \cite{Clauset:2009lr}, \cite{Valverde:2007fk}. We measure the greatest discrepancy between the empirical and the expected distribution (D-statistic), as well as the sum of squares of the deviations between the two distributions. Table~\ref{table1} summarizes the model parameters and the results. Note that the performance of other (perhaps simpler) models degrades when the entire range is considered.

\section{Discussion}

The proposed model generates extended power-law distributions from consecutive achievements of stable strength distributions $\mathcal{S}_k^*$. Although it does not pretend to empirically validate real-world mechanisms behind citation networks, the model may be of interest in the following context. First, it can be shown that the state $\mathcal{S}_k^*$ is a Nash, which implies that when a network reaches the equilibrium there is not any node that can gain by unilaterally rearranging weights to neighboring nodes (there are no incentives to change or establish new relationships). By focusing on the dynamics that drive the network to $\mathcal{S}_k^*$ we capture the coupling between different nodes, characterizing how relationships between any pair of nodes affects other nodes in the network. Second, the proposed strategies allow us to control the connectivity dynamics of nodes based on local attachment strategies (C1-5), allowing us to generate network substrates through distributed decision-making. Finally, the ability to control the rate at which attachment strategies lead to the scaling behavior allows us to obtain non-negligible clustering coefficients for large networks.

We focused on two types of network incentives: $(i)$ Longevity rewards nodes that have been part of the network for a long time (they have the ability to acquire more weight compared to recently added ones); $(ii)$ Fitness rewards nodes that are highly competent (they are more suitable to compete and maintain weights). Modeling nodes with varying fitness allows ``latecomers'' to overcome nodes that have been in the network for longer generations.


Following similar ideas as in Theorems 1 and 2, the proposed framework can be extended to generate exponential strength distributions. In particular, if we consider the gain function of the general from $g_i(s_i)=\frac{1}{s_i}\ln\left \vert \frac{1}{n_i}+\kappa \right \vert$ where $\kappa >  0$, the proposed strategies lead to weighted networks with $P[s_i > \omega] \sim e^{-\omega}$. A mathematical framework that allows us to generate various strength distributions for different domain intervals provides an important direction for future research. 

\section{Appendix}

\begin{proof}[Proof of Theorem 1]

First, we define a metric $\rho(s,\mathcal{S}^*_k)$ on the distribution of the strength and a Lyapunov function $V(s)$. We then show that for the choices of  $\rho(s,\mathcal{S}^*_k)$ and $V(s)$, $V(s)=0$ for $s \in \mathcal{S}^*_k$ and there exist two positive constant $c_1$ and $c_2$ such that $c_1\rho(s,\mathcal{S}^*_k) \le V(s) \le c_2\rho(s,\mathcal{S}^*_k)$ for all $s \in \mathcal{S}_k$. Finally, to prove asymptotic stability of $\mathcal{S}^*_k$ we show that for any initial distribution $s(0) \notin \mathcal{S}^*_k$ and any class of rewiring strategies that satisfy C1-3, i.e., for all $E^{1}_t$ such that $E^{1}_tE^{1} \in E^{1}(s(0))$, the functional $V(S(s(0),E^{1}_t,t)) \to 0$ as $t \to \infty$ for any fixed generation $k$.

Let $s'=[s'_{1}, \ldots , s'_{N_k}]^\top$ and choose
\begin{equation}\label{e7}
\rho(s,\mathcal{S}^*_k)=\inf \{ \max_{i}  \vert s_{i}-s'_{i} \vert \colon s' \in \mathcal{S}^*_k \}
\end{equation} and
\begin{equation}\label{e8}
V(s)=\max_{i} \{g_i(s_{i})\}-\frac{1}{N_k}\sum_{j \in \mathcal{H}_k}g_j(s_{j})
\end{equation}
Note that for $s \in \mathcal{S}^*_k$, $V(s)=0$ since $g_i(s_i)=g_j(s_j)$ for all $i, j \in  \mathcal{H}_k$ and $\rho(s,\mathcal{S}^*_k)=0$. To show that $V(s)$ is bounded from below by a class $\mathcal{K}$ function $c_1\rho(s,\mathcal{S}^*_k)$, note that according to eq.~\eqref{e1} for all $s_{i} \in (0,\infty)$ and all $i \in \mathcal{H}_k$, it must be the case that for any $s \notin \mathcal{S}^*_k$ and $s' \in \mathcal{S}^*_k$, there is some node $i \in  \mathcal{H}_k$ such that $s_{i} \ne s'_{i}$. Let $\underbar{$b$}=\min_i\{b_i\}$ and $c_1>0$ be a constant such that
\begin{equation}\label{e9}
\frac{g_i(s_{i})-g_i(s'_{i})}{s_{i}-s'_{i}} \le -b_i \le - \underbar{$b$} < - \frac{\underbar{$b$}}{N_k}= -c_1 < 0
\end{equation}
Since eq.~\eqref{e9} applies for any $i \in \mathcal{H}_k$ such that $s_{i} \ne s'_{i}$, it must apply for some node
\begin{equation}\label{e10}
r=\arg\max\{\vert s_{i}-s'_{i}\vert: s_{i} \ne s'_{i}\}
\end{equation}
Using the definition of $\rho(s,\mathcal{S}^*_k)$ and applying eq.~\eqref{e9} to node $r$ yields
\begin{eqnarray}\label{e11}
\underbar{$b$}\rho(s,\mathcal{S}^*_k) &\le& \underbar{$b$}\max\{ \vert s_{i}-s'_{i} \vert: s_{i} \ne s'_{i}\} \nonumber \\
&=&\underbar{$b$} \vert s_{r}-s'_{r}\vert \le \vert g_r(s_{r})-g_r(s'_{r})\vert
\end{eqnarray} 
Note that for any strength distribution $s \ne \mathcal{S}^*_k$ and $s' \in\mathcal{S}^*_k$, one of the following must be true: In the first case, if $g_r(s_{r})-g_r(s'_{r})>0$, i.e., if node $r$ needs more weight from its neighbors to achieve the desired state, then there must exist some other node $u$ such that
\begin{equation}\label{e12}
g_{u}(s_{u})-g_{u}(s'_{u}) < 0
\end{equation} 
In other words, there must exist another node $u$ that needs to weaken its relationship to neighboring nodes to achieve $\mathit{its}$ desired state $s'_{u}$. Because, $g_{u}(s'_{u})=g_r(s'_r)$ and $g_r(s'_r)>g_{u}(s_{u})$
\begin{equation}\label{e14}
g_r(s_r)-g_r(s'_r) < g_r(s_r)-g_{u}(s_{u}) \nonumber \\
\le \max_i \{g_i(s_i)\} - \min_i \{g_i(s_i)\}
\end{equation}
Similarly, if $g_r(s_r)-g_r(s'_r)<0$, i.e., if node $r$ needs to weaken its relationship to neighboring nodes to achieve the desired state, then there must also exist some other node $u$ such that
\begin{eqnarray*}\label{e13}
g_{u}(s_{u})-g_{u}(s'_{u}) > 0 
\end{eqnarray*} 
In other words, there must exist another node $u$ that needs to strengthen its relationships to achieve $\mathit{its}$ desired state $s'_{u}$. Because $g_{u}(s'_{u})=g_r(s'_r)$
\begin{eqnarray*}\label{e14}
0<g_r(s'_r)-g_r(s_r)&=&g_{u*}(s'_{u})-g_r(s_r) \\ \nonumber
&\le& g_{u}(s_{u})-g_r(s_r) \nonumber \\
&\le& \max_{i}\{g_i(s_i)\}-\min_{i}\{g_i(s_i)\}
\end{eqnarray*}
Thus, eq.~\eqref{e11} can be bounded from above by
\begin{eqnarray}\label{e15}
\underbar{$b$}\rho(s,\mathcal{S}^*_k) \le \max_{i}\{g_i(s_i)\}-\min_{i}\{g_i(s_i)\} 
\end{eqnarray} 
Next, note that
\begin{eqnarray*}\label{e16}
V(s) &\ge& \max_{i}{g_i(s_i)} -\frac{1}{N_k}\left[\min_{i}\{g_i(s_i)\}+(N_k -1)\max_{i}\{g_i(s_i)\}\right] \\
&\ge& \frac{1}{N_k}\left[\max_{i}\{g_i(s_i)\}-\min_{i}\{g_i(s_i)\}\right]
\end{eqnarray*} Using eq.~\eqref{e15} we get
\begin{equation}\label{e17}
\frac{\underbar{$b$}}{N_k}\rho(s,\mathcal{S}^*_k) \le \frac{1}{N_k}\left[\max_{i}\{g_i(s_i)\}-\min_{i}\{g_i(s_i)\}\right] \le V(s)
\end{equation}
Thus, $V(s) \ge c_1\rho(s,\mathcal{S}^*_k)$ for all $s \in \mathcal{S}_k$.

Next, we will show that there exist a constant $c_2$ such that $V(s) \le c_2 \rho(s,\mathcal{S}^*_k)$ for all $s \in \mathcal{S}_k$. Let $\bar{a}=\max_i\{a_i\}$. Recall that for all $s \notin \mathcal{S}^*_k$ and $s' \in \mathcal{S}^*_k$, $\max\{ \vert s_i-s'_i\vert\} > 0$. Note also that if $r=\arg\max_i\{g_i(s_i)\}$, then according to eq.~\eqref{e1}
\begin{equation}\label{e18}
0 \le \left\vert \frac{\max_i\{g_i(s_i)\}-g_{r}(s'_{r})}{\max\{ \vert s_i-s'_i\vert\}} \right\vert \le \bar{a}
\end{equation}
and similarly, if $u=\arg\min_i\{g_i(s_i)\}$, then
\begin{equation}\label{e19}
0 \le \left\vert \frac{g_u(s'_u)-\min_i\{g_i(s_i)\}}{\max\{ \vert s_i-s'_i\vert\}} \right\vert \le \bar{a}
\end{equation}
By adding eq.~\eqref{e18} and eq.~\eqref{e19} get
\small
\begin{eqnarray*}\label{e20}
2\bar{a} &\ge& \frac{\vert \max_i\{g_i(s_i)\}-g_{r}(s'_{r})\vert+\vert g_u(s'_u)-\min_i\{g_i(s_i)\} \vert}{\max\{ \vert s_i-s'_i\vert\}} \\
&\ge& \frac{\vert \max_i\{g_i(s_i)\}-g_{r}(s'_{r})+ g_u(s'_u)-\min_i\{g_i(s_i)\} \vert}{\max\{ \vert s_i-s'_i\vert\}}
\end{eqnarray*} \normalsize
Since $s' \in \mathcal{S}^*_k$, $g_u(s'_u)=g_{r}(s'_{r})$ and\begin{eqnarray*}\label{e21}
\frac{  \max_i\{g_i(s_i)\}-\min_i\{g_i(s_i)\}  }{ \max\{ \vert s_i-s'_i\vert\}} \le 2\bar{a}
\end{eqnarray*} 
Moreover
\begin{eqnarray*}\label{e22}
V(s) &\le& \max_i\{g_i(s_i)\}-\frac{1}{N_k}(N_k\min_i\{g_i(s_i)\}) \\
&\le& \max_i\{g_i(s_i)\}-\min_i\{g_i(s_i)\}
\end{eqnarray*}
Hence, if $c_2=2\bar{a}$
\begin{equation}\label{e23}
V(s) \le c_2 \max \{ \vert s_i-s'_i\vert\} 
\end{equation}
Since eq.~\eqref{e23} applies to any $s' \in \mathcal{S}^*_k$ and $s \notin \mathcal{S}^*_k$ and according to the definition of $\rho(s,\mathcal{S}^*_k)$
\begin{equation}\label{e24}
V(s) \le c_2 \inf \{\max_i \{ \vert s_i-s'_i\vert\}: s' \in \mathcal{S}^*_k \} = c_2\rho(s,\mathcal{S}^*_k)
\end{equation}
Thus, $V(s) \le c_2\rho(s,\mathcal{S}^*_k)$ for all $s \in \mathcal{S}_k$. 

Next, in order to show that $\mathcal{S}^*_k$ is globally asymptotically stable, we must show that for all $s(0) \notin \mathcal{S}^*_k$ and all $E^{1}_t$ such that $E^{1}_tE^{1} \in E^{1}(s(0))$,
\begin{equation}\label{e25}
V(S(s(0),E^{1}_t,t)) \to 0 \; \mbox{as} \; t \to \infty
\end{equation}
(i.e., $V \to 0$ along all possible motions of the system). This part of the proof is similar to the proof of Theorem 3.4 in \cite{Finke:2009ly}. If $s(t) \notin\mathcal{S}^*_k$, then there must exist some node $r \in \mathcal{H}_k$ with the highest gain among all nodes (there might actually be more than one). There must also exist another node $u \in \mathcal{H}_k$ such that $(u,r) \in {\mathcal{W}_k}$ and $g_u(s_u(t))<g_{r}(s_{r}(t))$. Because of the restrictions imposed by $E^1$, we know that events of type 1 are guaranteed to occur infinitely often. According to condition C2, when each event of type 1 $e_1 (t)$ occurs, the gain of node $r$ is guaranteed to decrease by a fixed fraction $\gamma \in (0,1)$ of $g_{r}(s_{r}(t))-g_u(s_u(t))$. Hence, if $e_{\mu(u)}^{\sigma(u)} \in e_1(t)$, then~$g_{r}(s_{r}(t+1)) < g_{r}(s_{r}(t))$. Regardless of how many nodes with the highest gain there are, since there are only a finite number of nodes in the network $\mathcal{G}_k$, it is inevitable that eventually the highest gain must decrease. Note that according to condition C3 no node can increase its gain beyond the gain of the highest nodes by weakening its relation from neighboring nodes. In other words, $\max_i\{g_i(s_i(t))\}$ must eventually decrease as long as $s(t) \notin \mathcal{S}^*_k$. Note also that since $\sum_{i \in \mathcal{H}_k}g_i(s_i(t))>0$, the Lyapunov function can be bounded by $0 \le V(s(t)) < \max_i\{g_i(s_i(t))\}$. Hence, for every $t \ge 0$, there exists $t'>t$ such that $V(s(t'))>V(s(t'+1))$ as long as $s(t') \notin \mathcal{S}^*_k$ so that $V(S(s(0),E^{1}_t,t)) \to 0$ as $t \to \infty$, and $\mathcal{S}^*_k$ has a region of asymptotic stability equal to $\mathcal{S}_k$.
\end{proof}

\begin{proof}[Proof of Lemma 1]

We show that $\forall i \in \mathcal{H}_k$ the value of $C_{k}$ converges as $k \to \infty$. Let $s(\tau^{+}_{k}) \in \mathcal{S}^*_{k}$. Using Theorem 1 we know
\begin{eqnarray*}
C_k = \frac{1}{N_k}\sum_{i \in \mathcal{H}_k}g_i (s_i)=\frac{1}{n_1 +k}\sum_{i \in \mathcal{H}_k}\frac{1}{s_i - s_0}\left ( \frac{k}{k_i}\right)^{\beta} 
\end{eqnarray*} 
Following assumption A2 for each generation $k$, $\forall i \in \mathcal{H}_k$ then $s_i - s_0 \ge \epsilon > 0$, so we have
\begin{eqnarray*}
\lim_{k \to \infty} C_k &\le& \lim_{k \to \infty} \frac{1}{\epsilon(n_1 +k)}\sum_{i \in \mathcal{H}_k}\left ( \frac{k}{k_i}\right)^{\beta} \\
&=& \frac{1}{\epsilon(1-\beta)}
\end{eqnarray*} 
Next, consider the difference in average gain between two consecutive generations
\begin{eqnarray*}
C_k-C_{k-1}&=&\frac{1}{n_1 +k}\sum_{i \in \mathcal{H}_k}\frac{1}{s_i -s_0}\left ( \frac{k}{k_i}\right)^{\beta}-\frac{1}{n_1 +k-1}\sum_{i \in \mathcal{H}_{k-1}}\frac{1}{s_i -s_0}\left ( \frac{k-1}{k_i}\right)^{\beta} \\
&=&\frac{1}{n_1 + k}\sum_{i \in \mathcal{H}_k}\frac{1}{s_i -s_0}\left(\frac{k}{k_i}\right)^{\beta}-\frac{1}{n_1 + k-1}\sum_{i \in \mathcal{H}_k}\frac{1}{s_i -s_0}\left(\frac{k-1}{k_i}\right)^{\beta} \\ 
&+& \frac{1}{n_1+k-1}\frac{1}{s_{N_k}(\tau^-_{k+1})-s_0}\left(\frac{k-1}{k}\right)^{\beta}\\
&>&\frac{1}{\epsilon(n_1+k)}\left(\frac{k-1}{k}\right)^{\beta}+\frac{1}{\epsilon(n_1 + k)} \left ( \sum_{i \in \mathcal{H}_k}\left(\frac{k}{k_i}\right)^{\beta}-\sum_{i \in \mathcal{H}_k}\left(\frac{k-1}{k_i}\right)^{\beta} \right ) \\
\mbox{Moreover,}  \\
C_k-C_{k-1}&<&\frac{1}{\epsilon(n_1+k-1)}\left(\frac{k-1}{k}\right)^{\beta}+\frac{1}{\epsilon(n_1 + k-1)} \left ( \sum_{i \in \mathcal{H}_k}\left(\frac{k}{k_i}\right)^{\beta}-\sum_{i \in \mathcal{H}_k}\left(\frac{k-1}{k_i}\right)^{\beta} \right ) \\
&=&\frac{1}{\epsilon(n_1+k-1)} \left ( \left(\frac{k-1}{k}\right)^{\beta} + \sum_{i \in \mathcal{H}_k}\left(\frac{k^\beta-(k-1)^{\beta}}{k_i^\beta}\right) \right )
\end{eqnarray*}
Because $\lim_{k \to \infty}\left(\frac{k}{k-1} \right)^{-\beta}=1$ and $\lim_{k \to \infty} \sum_{i \in \mathcal{H}_k}\frac{k^\beta-(k-1)^{\beta}}{k_i^\beta}=\frac{\beta}{1-\beta}$ 
\begin{eqnarray*}
\lim_{k \to \infty} C_k-C_{k-1}=0
\end{eqnarray*}
Let $C_{\infty}=\lim_{k \to \infty}C_k$. Because the weight added at the start of each generation is constant and $\frac{1}{s_i (\tau^+_k)-s_0}\left(\frac{k}{k_i}\right)^{\beta}=\frac{1}{s_j (\tau^+_k)-s_0}\left(\frac{k}{k_j}\right)^{\beta}=C_k, \ \forall_{i,j} \in \mathcal{H}_k$ such that $s \in \mathcal{S}^*_k$, using eq.~\eqref{e6} and eq.~\eqref{e50} we know
\begin{eqnarray*}\label{e26}
m&=&P_k-P_{k-1} \\
&=&\sum_{i \in \mathcal{H}_k}(s_i(\tau^{+}_{k})-s_0)-\sum_{i \in \mathcal{H}_{k-1}}(s_i(\tau^{+}_{k-1})-s_0) \\
m+s_0&=&\sum_{i \in \mathcal{H}_k}\frac{1}{C_{k}}\left(\frac{k_i}{k}\right)^{-\beta}-\sum_{i \in \mathcal{H}_{k-1}}\frac{1}{C_{k-1}}\left(\frac{k_i}{k-1}\right)^{-\beta} \\
&=&\frac{1}{C_{k}}\sum_{i \in \mathcal{H}_k}\left(\frac{k_i}{k}\right)^{-\beta}-\frac{1}{C_{k-1}}\sum_{i \in \mathcal{H}_{k-1}}\left(\frac{k_i}{k-1}\right)^{-\beta} \\
&=&\frac{1}{C_{k-1}}\left(\frac{k}{k-1}\right)^{-\beta}+\frac{1}{C_{k}}\sum_{i \in \mathcal{H}_{k}}\left(\frac{k^\beta-(k-1)^{\beta}}{k^\beta_i}\right)
\end{eqnarray*} 
letting $k \to \infty$, yields
\begin{eqnarray*}
m+s_0&=& \lim_{k \to \infty} \left( \frac{1}{C_{k-1}}\left(\frac{k}{k-1}\right)^{-\beta}+\frac{1}{C_{k}}\sum_{i \in \mathcal{H}_{k}}\left(\frac{k^\beta-(k-1)^{\beta}}{k^\beta_i}\right)
\right ) \\
&=& \lim_{k \to \infty} \frac{1}{C_{k-1}}\left(\frac{k}{k-1}\right)^{-\beta} + \lim_{k \to \infty}\frac{1}{C_{k}}\sum_{i \in \mathcal{H}_{k}}\left(\frac{k^\beta-(k-1)^{\beta}}{k^\beta_i}\right) \\
&=&\lim_{k \to \infty}\frac{1}{C_{k-1}}+\left(\frac{\beta}{1-\beta}\right)\lim_{k \to \infty}\frac{1}{C_{k}}=\frac{1}{C_{\infty}}\left( 1 + \frac{\beta}{1-\beta} \right)=\frac{1}{C_{\infty}(1-\beta)}
\end{eqnarray*}
As $k \to \infty$, the average gain is given by
\begin{eqnarray}\label{e28}
C_{\infty}=\frac{1}{(m+s_0)(1-\beta)}
\end{eqnarray}
\end{proof}

\begin{proof}[Proof of Theorem 2]

Because of the restrictions imposed by $E^{2}_kE^{2}$, we know that events of type 2 occur infinitely often (i.e., new nodes are persistently added to the network). Using Theorem 1 and according to eq.~\eqref{e3} and eq.~\eqref{e6} $\forall i \in \mathcal{H}_k$
\begin{equation*}\label{e35}
\frac{1}{s_i(\tau^+_k)-s_0}\left (\frac{n_i} {1} \right )^{- \beta}=C_{k}\end{equation*} 
Because $n_i=\frac{k_i}{k}$ we know that 
\begin{equation*}\label{e36}
s_{i}(\tau^+_{k}) - s_0=\frac{1}{C_{k}} \left ( 1\frac{k}{k_i} \right )^{\beta} 
\end{equation*} 
which indicates that the strength of any node $i$ follows a power-law distribution over generations. Following similar ideas as in \cite{Valverde:2002kx}, the probability that a node has strength $s_{i} - s_0$ smaller than $\omega$ is
\begin{eqnarray}\label{e37}
P[s_{i} -s_0< \omega]=P\left [\frac{1}{C_{k}} \left (1 \frac{k}{k_i} \right)^\beta - s_0 < \omega \right ]=P\left[ \frac{1}{k_i} < \frac{C_{k}^{1/\beta} (\omega + s_0)^{1/\beta}}{k} \right] \nonumber
\end{eqnarray} 
where $k_i$ (i.e., the generation at which every node is added to the network) follows a probability density function $k_i \sim U(1,k)$. 
To characterize the emergence of scaling behavior note that 
\begin{eqnarray*}
P\left[\frac{1}{k_i} < \frac{C_{k}^{1/\beta}(\omega + s_0)^{1/\beta}}{k}\right ]&=&P\left[k_i > kC_{k}^{-1/\beta}(\omega + s_0)^{-1/\beta}\right ] \\
&=&1-C_{k}^{-1/\beta}(\omega + s_0)^{-1/\beta}
\end{eqnarray*}
The strength distribution is given by 
\begin{eqnarray*}
p_{\omega}=\frac{\partial P[s_{i} - s_0<\omega]}{\partial \omega}
 &=& \frac{\partial P\left[k_i > kC_{k}^{-1/\beta}(\omega + s_0)^{-1/\beta}\right]}{\partial \omega} \\ 
&=&\frac{1}{\beta}C^{-\frac{1}{\beta}}_{k}(\omega + s_0)^{-\frac{1}{\beta}-1} 
\end{eqnarray*} 
If we let $k \to \infty$, then
\begin{eqnarray}\label{e40}
p_{\omega}=\frac{1}{\beta}C^{-\frac{1}{\beta}}_{\infty}(\omega + s_0)^{-\alpha} 
\end{eqnarray} which leads to an extended power-law distribution with ${\alpha=\frac{1}{\beta}+1}$.

Following similar ideas as in \cite{Newman:2005lq} we now characterize the value at which the power-law emerges. The general expression for a power-law exhibits
\begin{eqnarray}\label{e44}
p_{\omega}=C\omega^{-\alpha} \nonumber
\end{eqnarray}
where the normalization constant $C$ can be expressed as $C=(\alpha-1)\hat{s}^{\alpha-1}$, then\begin{eqnarray}\label{e45}
\frac{1}{\beta}C^{-\frac{1}{\beta}}_{\infty}&=&(\alpha-1)\hat{s}^{\alpha-1} \nonumber \\
C_{\infty}&=&\hat{s}^{-1}
\end{eqnarray} 
Using eq.~\eqref{e28}, eq.~\eqref{e40} and eq.~\eqref{e45} we have
\begin{eqnarray}\label{e46}
\hat{s}=(m + s_0)(1-\beta)=\frac{\alpha-2}{\alpha-1}(m + s_0)
\end{eqnarray}

\end{proof}

\bibliographystyle{ieeetr} 
\bibliography{Anteproyecto}

\end{document}